\documentclass[twocolumn,twoside,superscriptaddress,prl,aps,showpacs]{revtex4}
\usepackage{amsmath,mathrsfs,amsbsy,color,graphicx,bm,amsthm,amsfonts}
\usepackage{units}
\usepackage{bbm}
\usepackage{times}
\usepackage{dcolumn}
\usepackage{mathrsfs}
\usepackage{amsmath,amssymb,epsfig,float}

\begin{document}

\title{Satellite-based quantum steering under the influence of spacetime curvature of the Earth }

\author{Tonghua Liu$^{1}$,  Jiliang Jing$^{1}$, and Jieci Wang$^{1}$\footnote{Email: jieciwang@hunnu.edu.cn}}
\affiliation{Department of Physics, and Synergetic Innovation Center for Quantum Effects \\
and Applications,
 Hunan Normal University, Changsha, Hunan 410081, China
}

\begin{abstract}
Spacetime curvature of the Earth  deforms wavepackets of photons sent from the Earth to satellites, thus influencing the  quantum state of light. We show that Gaussian steering of photon pairs, which are initially prepared in a two-mode squeezed state, is affected by the curved spacetime background of the Earth.  We demonstrate that quantum steerability of the state increases for a specific range of height $h$ and then  gradually approaches a finite value with further increasing height of the satellite's orbit. Comparing with the peak frequency parameter, the Gaussian steering changes more for different squeezing parameters, while  the gravitational
frequency effect leads to quantum steering asymmetry between the photon pairs. In addition, we find that the influence of spacetime curvature on the steering in the Kerr spacetime is very different from the non-rotating case because special relativistic effects are involved.
\end{abstract}
\maketitle

\section{Introduction}

Einstein-Podolsky-Rosen (EPR) steering, which describes how the state of one subsystem in an entangled pair is manipulated by local measurements performed on the other part, was proposed by Schr\"{o}dinger in 1935 \cite{E.P.C, E.P.C1}. The EPR steering differs from both   entanglement and  the Bell nonlocality, because it has inherent asymmetric features. Therefore, the EPR steering has potential applications in the one-side device-independent quantum key distribution \cite{CBEG}. Recently, quantum steering and its asymmetry have been
theoretically studied \cite{HMW,Skrzypczyk, Kocsis, Adesso2015, steering3, steering4,MWZZ,steering5,SLCC} and experimentally demonstrated \cite{Walborn,steering2,Bowles,VHTE,VHTSS,Handchen,BWSR,SKMJ,DJSS,TEVH,Xiao2017,SWNW} in different quantum systems.  However, little is known about behaviors of quantum steering in  relativistic settings. Most recently, Navascues and Perez-Garcia studied quantum steering between space-like separated parties in the frame of algebraic quantum field theory \cite{steeringrqi}. In addition, more attention has been given to the dynamics of quantum steering under the influence of the dynamical Casimir effect  \cite{steeringrqi1},  the Hawking radiation  \cite{steeringrqi2}, and  relativistic motions \cite{steeringrqi3}.

Since a realistic quantum system cannot be  prepared and transmitted in a curved
spacetime without any gravitational and relativistic effects,
the study of quantum steerability in a relativistic framework is necessary. Such studies are of practical and  fundamental importance to understand the influence of gravitational effects on the steerability-type quantum resource when the parties involved are located at large distances in the curved space time \cite{DEBA,DEBT,satellite1,satellite2,kerr}. It has been shown that the
curved background spacetime of the Earth  affects the running of quantum clocks \cite{Alclock},  is employed as witnesses of general
relativistic proper time in laser interferometric \cite{Zych}, and  influences the implementation of quantum metrology \cite{MADE, MADE2} in satellite-based setups. Furthermore, Kish and Ralph found that there would be inevitable losses of quantum resources
in the estimation of the Schwarzschild radius \cite{SPK}.
We studied how the curved background spacetime of the Earth  influences the satellite-based quantum clock synchronization \cite{wangsyn}.  Most recently, an experimental test of photonic entanglement in an accelerated setting  was realized \cite{RQI8}, where a genuine quantum state of entangled photon pairs was exposed to different accelerations.

In this work, we present a quantitative investigation of Gaussian
quantum steerability for correlated photon pairs which are initially prepared in a two-mode squeezed state in the
curved background spacetime of the Earth. We assume that one of entangled photons is sent to Alice (at the Earth station) and the other propagates to Bob (at the satellite). During this propagation, the photons' wave-packet will be deformed by the curved background spacetime of the Earth, and these deformations effects  on the quantum state of the photons can be modeled as a lossy quantum channel \cite{MANI, wangsyn}. Since the initial state is Gaussian and the transformations involved are linear and unitary, we can restrict our state to the  Gaussian scenarios and employ the covariance matrix formalism. We  calculate the
Gaussian quantum steering from Alice to Bob, which quantifies to what
extent Bob's mode can be steered by Alice's measurements. We also discuss Gaussian quantum steering from Bob to Alice to verify the asymmetric property of steerability in the curved spacetime.

This work is organized as follows. In section II, we introduce the quantum field theory of a massless uncharged bosonic field which propagates from the Earth to a satellite. In section III, we briefly introduce the definition and measure of the bipartite Gaussian quantum steering. In section IV, we show a scheme to test quantum steering between the Earth and satellites and study the behaviors of quantum steering in the curved spacetime.  The last section is devoted to a brief summary. Throughout the whole paper we employ the natural units $G = c =\hbar= 1$.

\section{Light wave-packets propagating in the curved space-time \label{tools}}
In this section we will describe the propagation of photons from the Earth to satellites under the influence  of the Earth's gravity \cite{DEBT}. The Earth's spacetime can be approximately described by the Kerr metric \cite{Visser}. For the sake of simplicity, our work will be constrained to the equatorial plane $\theta=\frac{\pi}{2}$. The reduced metric in Boyer-Lindquist coordinates $(t,r,\phi)$  reads \cite{Visser}
\begin{align}\label{metric}
ds^2=&\, -\Big(1-\frac{2M}{r} \Big)dt^2+\frac{1}{\Delta}dr^2 \nonumber \\
&\,+\Big(r^2+a^2+\frac{2Ma^2}{r}\Big) d\phi^2 - \frac{4Ma}{r} dt \, d\phi, \\
\Delta=&\,1-\frac{2M}{r}+\frac{a^2}{r^2},
\end{align}
where $M$, $r$, $J$, $a=\frac{J}{M}$ are the mass, radius, angular momentum and Kerr parameter of the Earth, respectively.

A photon is sent from Alice on Earth's surface to Bob at time $\tau_A$, Bob will receive this photon at $\tau_B=\Delta\tau+\sqrt{f(r_B)/f(r_A)}\tau_A$ in his own reference frame, where $f(r_A)=1-\frac{r_S}{r_A}$ and  $f(r_B)=1-\frac{r_S}{r_B}$. Here $r_S=2M$ is the Schwarzschild radius of
the Earth and $\Delta\tau$ is the propagation time  of the light from the Earth to the satellite by taking curved effects of the Earth into account. In general, a photon can be modeled by a wave packet of excitations of a
massless bosonic field with a distribution $F^{(K)}_{\Omega_{K,0}}$ of mode frequency $\Omega_{K}$ and peaked at $\Omega_{K,0}$ \cite{ULMQ,TGDT}, where $K=A,B$ denote the modes in Alice's or Bob's reference frames, respectively. The annihilation operator of a
photon for an observer far from  Alice or Bob takes the form
\begin{equation}
a_{\Omega_{K,0}}(t_K)=\int_0^{+\infty}d\Omega_K e^{-i\Omega_K t_K}F^{(K)}_{\Omega_{K,0}}(\Omega_K)a_{\Omega_K}.
\label{wave}
\end{equation}
Alice's and Bob's operators in Eq. (\ref{wave}) can be used to describe the same optical mode in  different altitudes. By considering the curved spacetime  of the Earth,
the wave packet received is modified. The relation between $a_{\Omega_A}$ and $a_{\Omega_B}$ was discussed in \cite{DEBT,DEBA,wangsyn}, and can be used to calculate the relation between the frequency distributions $F^{(K)}_{\Omega_{K,0}}$ of the photons before and after the propagation \cite{DEBT,DEBA,wangsyn}
\begin{eqnarray}
F^{(B)}_{\Omega_{B,0}}(\Omega_B)=\sqrt[4]{\frac{f(r_B)}{f(r_A)}}F^{(A)}_{\Omega_{A,0}}\left(\sqrt{\frac{f(r_B)}{f(r_A)}}\Omega_B\right).\label{wave:packet:relation}
\label{fab}
\end{eqnarray}
From Eq. (\ref{fab}), we can see that the effect induced by the
curved spacetime  of the Earth cannot be simply corrected by a linear shift of frequencies. Therefore, it may be challenging to compensate the transformation induced by the curvature in realistic implementations.

Indeed, such a nonlinear gravitational effect is found to influence the fidelity of the quantum channel between Alice and Bob \cite{DEBT,DEBA,wangsyn}. It is always possible to decompose the mode $\bar{a}^{\prime}$ received by Bob in terms of the mode $a^{\prime}$ prepared by Alice  and an orthogonal mode $a_{\bot}^{\prime}$ (i.e. $[a^{\prime},a_{\bot}^{\prime\dagger}]=0$) \cite{PPRW}
\begin{eqnarray}
\bar{a}^{\prime}=\Theta a^{\prime}+\sqrt{1-\Theta^2}a_{\bot}^{\prime},\label{mode:decomposition}
\end{eqnarray}
where $\Theta$ is the wave packet overlap between the distributions $F^{(B)}_{\Omega_{B,0}}(\Omega_B)$ and $F^{(A)}_{\Omega_{A,0}}(\Omega_B)$,
\begin{eqnarray}
\Theta:=\int_0^{+\infty}d\Omega_B\,F^{(B)\star}_{\Omega_{B,0}}(\Omega_B)F^{(A)}_{\Omega_{A,0}}(\Omega_B),\label{single:photon:fidelity}
\end{eqnarray}
and  we have $\Theta=1$ for a perfect channel. From this expression we can see that  the spacetime curvature of the Earth would affect the  the fidelity $\mathcal{F}=|\Theta|^2$  as well as  the quantum resource of EPR steering.

We assume that Alice employs a real normalized Gaussian wave packet
\begin{eqnarray}
F_{\Omega_0}(\Omega)=\frac{1}{\sqrt[4]{2\pi\sigma^2}}e^{-\frac{(\Omega-\Omega_0)^2}{4\sigma^2}}\label{Bobpacket},
\end{eqnarray}
with  wave packet width $\sigma$. In this case the overlap $\Theta$ is given by \eqref{single:photon:fidelity} where we have extended the domain of integration to all the real axis. We note that the integral should be performed over strictly positive frequencies. However, since $\Omega_0\gg \sigma$, it is possible
to include negative frequencies without affecting the value of $\Theta$. Using Eqs. \eqref{wave} and \eqref{Bobpacket} one finds that \cite{DEBT,DEBA,wangsyn}
\begin{eqnarray} \label{theta}
\Theta=\sqrt{\frac{2}{1+(1+\delta)^2}}\frac{1}{1+\delta}e^{-\frac{\delta^2\Omega_{B,0}^2}{4(1+(1+\delta)^2)\sigma^2}}\label{final:result},
\end{eqnarray}
where the new parameter $\delta$ quantifying the shifting is defined by
\begin{equation}
\delta=\sqrt[4]{\frac{f(r_A)}{f(r_B)}}-1=\sqrt{\frac{\Omega_B}{\Omega_A}}-1.
\end{equation}
The expression for $\frac{\Omega_B}{\Omega_A}$ in the equatorial plane of the Kerr spacetime has been shown in \cite{kerr}
\begin{equation}\label{aw}
\frac{\Omega_B}{\Omega_A}=\frac{1+\epsilon \frac{a}{r_B}\sqrt{\frac{M}{r_B}}}{C\sqrt{1-3\frac{M}{r_B}+
2\epsilon\frac{a}{r_B}\sqrt{\frac{M}{r_B}}}},
\end{equation}
where $C=[1-\frac{2M}{r_A}(1+2a {\omega})+\big(r^2_A+a^2-\frac{2Ma^2}{r_A}\big){\omega}^2]^{-\frac{1}{2}}$ is the normalization constant, $\omega$ is the Earth's equatorial angular velocity and $\epsilon=\pm1$ stand for the direct of orbits (i.e., when $\epsilon=+1$ for the satellite co-rotates with the Earth). In the Schwarzschild limit $a, \omega\rightarrow0$,  Eq. (\ref{aw}) coincides to the result found in \cite{DEBT}, which is
\begin{equation}
\frac{\Omega_B}{\Omega_A}=\sqrt{\frac{1-\frac{2M}{r_A}}{1-\frac{3M}{r_B}}}.
\end{equation}

Noticing that $(r_A \omega)^2>a\omega$, therefore we can retain second order terms in $r_A\omega$. Expanding Eq. (\ref{aw}) we
obtain the following perturbative expression for $\delta$. This perturbative result does not depend
on whether the Earth and the satellite are co-rotating or not
\begin{eqnarray}\label{bw}
\nonumber\delta&=&\delta_{Sch}+\delta_{rot}+\delta_h\\
\nonumber&=&\frac{1}{8}\frac{r_S}{r_A}\big(\frac{1-2\frac{h}{r_A}}{1+\frac{h}{r_A}} \big)-\frac{(r_A\omega)^2}{4}-\frac{(r_A\omega)^2}{4}\big(\frac{3}{4}\frac{r_S}{r_A}-\frac{4Ma}{\omega r_A^3}\big),
\end{eqnarray}
where $h=r_B-r_A$ is the height between Alice and Bob, $\delta_{Sch}$ is the first order Schwarzschild term, $\delta_{rot}$ is the lowest order rotation term and $\delta_h$ denotes all higher order correction terms. If the parameter $\delta=0$ (the satellite moves at the height $h\simeq\frac{r_A}{2}$),  we have $\Theta=1$. That is to say,  the received photons at this height will not experience any frequency shift, and the effects of gravity of
the Earth and the effects of special relativity completely compensates each other.

\section{Gaussian quantum steering}

In this section we briefly review the measurement of quantum steering for a  general two-mode Gaussian state $\rho_{AB}$. The character of a bipartite Gaussian state ${\rho _{AB}}$ can be described by its covariance matrix (CM)
\begin{equation}\label{CM}
\sigma_{AB} = \left( {\begin{array}{*{20}{c}}
   A & C  \\
   {{C^{\sf T}}} & B  \\
\end{array}} \right),
\end{equation} with elements ${\sigma _{ij}} = \text{Tr}\big[ {{{\{ {{{\hat R}_i},{{\hat R}_j}} \}}_ + }\ {\rho _{AB}}} \big]$. Here the submatrices $A$ and $B$ are the CMs correspoding to the reduced states of $A$'s and $B$'s subsystems, respectively. The \textit{bona fide} condition should be  satisfied for a physical CM, which is
\begin{equation}\label{bonafide}
{\sigma _{AB}} + i\,({\Omega _A} \oplus {\Omega _B}) \ge 0.
\end{equation}

Let us continue by giving the definition of steerability. For a bipartite state, it is steerable from $A$ to $B$ \textit{iff} it is \textit{not} possible for every pair of local observables $R_A \in \mathcal{M}_A$ on $A$ and $R_{B}$ (arbitrary) on $B$, with respective outcomes $r_A$ and $r_{B}$, to express the joint probability as   \cite{HMWS}
$P\left( {{r_A},{r_{B}}|{R_A},{R_{B}},{\rho _{AB}}} \right) = \sum\limits_\lambda  {{\wp_\lambda }} \, \wp\left( {{r_A}|{R_A},\lambda } \right)P\left( {{r_{B}}|{R_{B}},{\rho _\lambda }} \right)$. That is to say, there exists at least one measurement pair between $R_A$ and $R_{B}$ that can violate this expression when ${\wp_\lambda }$ is fixed across all measurements. Here ${\wp_\lambda }$ and $\wp \left( {{r_A}|{R_A},\lambda }\right)$ are arbitrary probability distributions and $P\left( {{r_{B}}|{R_{B}},{\rho _\lambda }} \right)$ is a probability distribution restricted to the extra condition of being evaluated on a quantum state $\rho_\lambda$.
It has been proven that a necessary and sufficient condition for Gaussian $A\to B$ steerability is \textit{iff} the condition
\begin{equation}\label{nonsteer}
{\sigma _{AB}} + i\,({0_A} \oplus {\Omega _B}) \ge 0,
\end{equation}
is violated  \cite{HMWS}. To quantify how much a bipartite  Gaussian state with CM $\sigma_{AB}$ is steerable (by Gaussian measurements on Alice's side),  the following quantity has been performed  \cite{IKAR}
\begin{equation}\label{GSAB}
{\cal G}^{A \to B}(\sigma_{AB}):=
\max\bigg\{0,\,-\sum_{j:\bar{\nu}^{B}_j<1} \ln(\bar{\nu}^{B}_j)\bigg\}\,,
\end{equation}
where $\bar{\nu}^{B}_j$ are the symplectic eigenvalues of the Schur complement
of $A$ in the covariance matrix $\sigma_{AB}$.
 By defining the Schur complement $\det{\sigma_{AB}} = \det{A} \det{M^{B}_{\sigma}}$ and employing the R\'enyi-$2$ entropy, Eq. (\ref{GSAB}) can be written as
\begin{eqnarray}\label{GSAC}
{\cal G}^{A \to B}(\sigma_{AB})& =&
\nonumber\mbox{$\max\big\{0,\, \frac12 \ln {\frac{\det A}{\det \sigma_{AB}}}\big\}$}\\
&=& \max\big\{0,\, {\cal S}(A) - {\cal S}(\sigma_{AB})\big\}\,, \label{GS1}
\end{eqnarray}
where the  R\'enyi-$2$ entropy ${\cal S}$  reads ${\cal S}(\sigma) = \frac12 \ln( \det \sigma)$ \cite{renyi}   for a Gaussian state with CM $\sigma$. However,  unlike quantum entanglement, quantum steering is asymmetric \cite{IKAR}. To obtain the measurement of Gaussian steering $B\rightarrow A$, one can swap the roles of $A$ and $B$ and get an expression like Eq. (\ref{GSAC}).

\section{The influence of gravitational effects on  quantum steerability and entanglement}

In this section we propose a scheme to test large distance quantum steering between the Earth and satellites and discuss how quantum steerability is affected by the curved spacetime  of the
Earth. Firstly, we consider a pair of entangled photons
which are initially prepared in a two-mode squeezed state with modes $b_1$ and $b_2$ at the  ground station. Then we send one photon with mode $b_1$ to Alice. The other photon in mode $b_2$ propagates from the Earth to the satellite and is received by Bob. Due to the curved background spacetime of the Earth, the wave packet of photon with mode $b_2$ is deformed. Finally, one can
test how the quantum state of Alice's photon is manipulated by local Gaussian measurements performed by Bob at the satellite and verifies the quantum steerability from $b_2$ to $b_1$, and vice versa.

Considering that Alice receives the mode $b_1$ and Bob receives the mode $b_2$ at different satellite orbits, we should take the curved spacetime  of the Earth into account. As discussed in \cite{DEBT,DEBA,wangsyn}, the influence of the Earth's  gravitational   effect can be modeled by a beam splitter with orthogonal modes $b_{1\bot}$ and $b_{2\bot}$. The covariance matrix of the initial state is given by
\begin{equation}\label{initialcov}
\Sigma^{b_1b_2b_{1\bot}b_{2\bot}}_0=\left(
\begin{array}{cc} \tilde\sigma(s) &0 \\ 0  &  I_4
\end{array}\right),
\end{equation}
where ${I}_4$ denotes the $4\times4$ identity matrix and $\tilde\sigma{(s)}$ is  the covariance matrix of the two-mode squeezed state
\begin{equation}
\tilde\sigma(s)=\left(
\begin{array}{cc} \cosh{(2s)}  {I}_2&\sinh{(2s)}\sigma_z \\ \sinh(2s)\sigma_z &\cosh{(2s)}  {I} _2
\end{array}\right),
\end{equation}
where $\sigma_z$ is Pauli matrix and $s$ is  the squeezing parameter. The effect induced by the
curved spacetime  of the Earth on Bob's mode $b_2$ can be model as a lossy channel, which is described by the transformation \cite{DEBT,DEBA,wangsyn}
\begin{eqnarray}
\bar{b}_2&=&\Theta_2\,b_2+\sqrt{{1-\Theta_2^2}}b_{2\bot},
\end{eqnarray}
while the mode $b_1$ received by Alice is unaffected because Alice stays at the ground station.
This process can be represented as a mixing (beam splitting ) of modes $b_1(b_2)$ and $b_{1\bot}(b_{2\bot})$. Therefore, for the entire state, the symplectic transformation can be encoded into the Bogoloiubov transformation
\begin{equation}
S=\left(
\begin{array}{cccc}
  {I}_2 &0&   0&0  \\ 0&\Theta_2  {I} _2 &0&\sqrt{
1-\Theta_2^2}  {I} _2\\  0 &0&  -{I}_2&0  \\ 0&\sqrt{
1-\Theta_2^2}  {I} _2 &0&-\Theta_2  {I} _2
\end{array}\right).\nonumber
\end{equation}
The final state  $\Sigma^{b_1b_2b_{1\bot}b_{2\bot}}$ after the transformation is  $\Sigma^{b_1b_2b_{1\bot}b_{2\bot}}=S\,\Sigma_0^{b_1b_2b_{1\bot}b_{2\bot}}\,S^{T}$.  Then we trace over the orthogonal modes $b_{1\bot},b_{2\bot}$ and obtain the covariance matrix $\Sigma^{b_1b_2}$ for the modes $b_1$ and $b_2$  after the propagation
\begin{equation}\Sigma^{b_1b_2}=\left(
\begin{array}{cc}\label{gst}
(1+2\sinh^2s ) {I}_2 &\sinh{(2s)}\,\Theta_2\,\sigma_z  \\ \sinh{(2s)}\,\Theta_2\,\sigma_z &(1+2\sinh^2s\,\Theta_2^2 )\, {I}_2
\end{array}\right).
\end{equation}

The form of the two-mode squeezed state under the influence of the effects of gravity of the Earth is given by Eq. (\ref{gst}). Then employing the measurement of Gaussian steering, we obtain an specific mathematic expression of the mode $b_1\rightarrow b_2$ Gaussian steering under the curved spacetime  of the Earth
\begin{eqnarray}\label{gaussian5}
{\cal G}^{b_1\to b_2} &=&
\mbox{$\max\big\{0,\, \ln {\frac{ 1+2\sinh^2(2s)}{1+2(1-
\Theta_2^2)\sinh^2s}}\big\}$}.
\end{eqnarray}
We notice that the wave packet overlap $\Theta$ in the above equation  is determined by the parameters $\delta$, $\sigma$ and $\Omega_{B,0}$.  Since the Schwarzschild radius of the Earth is $ r_s= 9$ mm, we have
$\delta\sim-\frac{1}{2}(\frac{r_s}{r_B}-\frac{r_s}{r_A}) \sim 10^{-10}$.  Here we consider a typical PDC source with a
wavelength of 598 nm (corresponding to the peak frequency $\Omega_{B,0}= 500$ THz) and   Gaussian bandwidth $\sigma=1$MHz [48, 49]. Under these constraints, $\delta\ll(\frac{\Omega_{B,0}}{\sigma})^2\ll1$ is
satisfied.
Therefore, the wave packet overlap $\Theta$ can be expand by the parameter $\delta$.  Then we obtain $\Theta \sim1-\frac{\delta^2\Omega_{B,0}^2}{8\sigma^2}$ by keeping the second order terms.  The Eq. (\ref{gaussian5}) has following form in the second order of perturbation for the parameter $\delta$
\begin{equation}\label{g0}
 {\cal G}^{b_1\to b_2}\simeq\max\{0,{\cal G}_0-\frac{\delta^2\Omega_{B,0}^2}{2\sigma^2}\sinh^2(s)\},
\end{equation}
where higher order contributions are neglected. To ensure the validity of  perturbative expansion, we should estimate the values of the last term in Eq. (\ref{g0}).   Considering $\frac{\delta^2\Omega^2_{B,0}}{2\sigma^2}\sim1.25\times10^{-7}$, we find that even if the value of the squeezing parameter is $s\ll7.6$ (corresponding to $\sinh^2(s)\ll10^6$), the perturbative expansion is valid. Therefore, we can safely prelimit the value of the squeezing parameter as $s<3$ hereafter. In the case of flat spacetime, this expression reduces to  ${\cal G}_0:=\ln{[1+2\sinh^2(2s)]}$.  As showed in Eq. (\ref{g0}), the Gaussian steering $b_1\to b_2$ not only depends on the squeezing parameter, the peak frequency, and the  Gaussian bandwidth, but also  the height of the orbiting satellite. This means that the curved spacetime  of the Earth will influence the $b_1\to b_2$ steerability because  the parameter $\delta$ contains the height $h$ of the satellite. It is clear  that $\delta$ approaches to a constant value when the height $h\rightarrow\infty$ and the squeezing parameter $s$ is a fixed value. Therefore, quantum steering ${\cal G}^{b_1\to b_2}$ also becomes a constant.

\begin{figure}[tbp]

\centering
\centerline{\includegraphics[width=7.0cm]{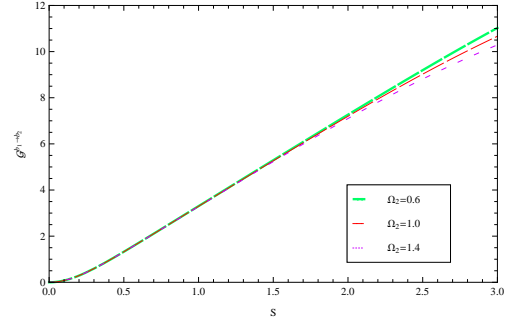}}
\caption{(Color online) The Gaussian steering ${\cal G}^{b_1\to b_2}$ of two-mode squeezed state as a function of the squeezing parameter $s$ for different peak frequencies, $\Omega_2=0.6$ (green dashed line), $\Omega_2=1$ (red dashed line) and $\Omega_2=1.4$ (violet dotted line), respectively. The orbit height of the satellite and the   Gaussian bandwidth are fixed as $h=20000$km and  $\sigma=1$.
}\label{f1}
\end{figure}

For convenience, we will work with dimensionless quantities by rescaling the peak frequency and the Gaussian bandwidth
\begin{equation}
\Omega \rightarrow \tilde{\Omega}\equiv\frac{\Omega}{\Omega_{B,0}}, \sigma \rightarrow \tilde{\sigma}\equiv\frac{\sigma}{\sigma_0},
\end{equation}
where $\Omega_{B,0}=500$THz and $\sigma_0=1$ MHz. For simplicity, we abbreviate  the dimensionless parameter $\tilde{\Omega}$ as $\Omega_2$ and abbreviate $\tilde{\sigma}$ as  $\sigma$, respectively.

In Fig. (1) we plot the Gaussian steering ${\cal G}_{b_1\to b_2}$ as a function of the squeezing parameter $s$
for the fixed orbit height $h=20000$ km and Gaussian bandwidth $\sigma=1$. We can see that quantum steering  monotonically increases with the increase of the squeezing parameter $s$. It is also shown that, comparing with the peak frequency parameter, the Gaussian steering $\mathcal{G}^{b_1\to b_2}$ changes more for different squeezing parameters, which indicates that the initial quantum resource plays a more important role in the quantum steering.

The Gaussian steering ${\cal G}_{b_1\to b_2}$ in terms of the orbit height $h$ and the   Gaussian bandwidth $\sigma$ for the fixed values $s=1$ and $\Omega_2=1$ has been shown in Fig. (2). We can see that the quantum steerability ${\cal G}_{b_1\to b_2}$ decreases with increasing  the Gaussian bandwidth $\sigma$. In addition, comparing with the squeezing parameter, the Gaussian steering is not easy to change with changing orbit height parameter and Gaussian bandwidth. This allows  us to choose appropriate physical parameters and perform more reliable quantum steering tasks between the Earth to a satellite.

\begin{figure}[tbp]
\centering
\includegraphics[height=2.3in, width=2.6in]{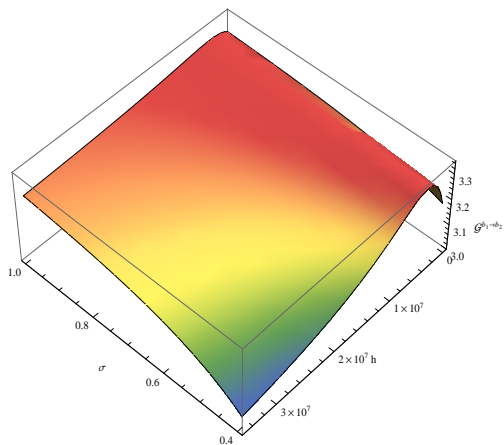}
\caption{(Color online) The Gaussian steering ${\cal G}_{b_1\to b_2}$ in terms of the orbit height $h$ and the   Gaussian bandwidth $\sigma$, for the fixed values $s=1$ and $\Omega_2=1$. }
\end{figure}

\begin{figure}[tbp]
\centerline{\includegraphics[width=7.7cm]{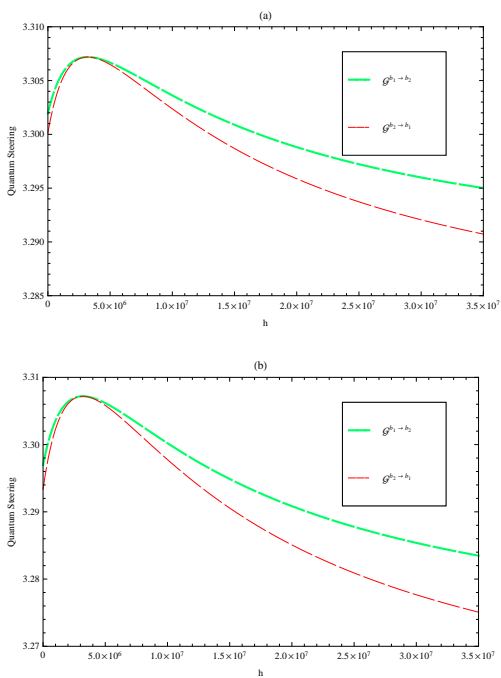}}
\caption{ (Color online). The Gaussian steering ${\cal G}^{b_1\to b_2}$ (green lines) and ${\cal G}^{b_2\to b_1}$ (orange lines) as functions of the height between Alice and Bob under the influence of the Earth's
gravity. Here the   Gaussian bandwidth of the initial state is fixed as $\sigma=1$, the dimensionless peak frequencies of the mode $b_2$ are fixed as (a) $\Omega_{2}=0.6$, (b) $\Omega_{2}=1$, and the squeezing parameter is $s=1$.
}\label{f1}
\end{figure}

One of the most distinguishable properties of quantum steering is its asymmetry, which has been recently experimentally demonstrated in flat spacetime \cite{VHTE,VHTSS}. To understanding this properties in the curved spacetime, we also calculate the steerability ${\cal G}^{b_2\to b_1}$, which is
\begin{eqnarray}
 {\cal G}^{b_2\to b_1} &=&
\mbox{$\max\big\{0,\,  \ln {\frac{ 1+2\sinh^2(2s)\Theta_2^2}{1+2(1-
\Theta_2^2)\sinh^2s }}\big\}$},
\end{eqnarray}
Similarly, this equation can be rewritten in its perturbative expansion
form as
\begin{eqnarray}
 {\cal G}^{b_2\to b_1}\simeq\max{\{0,\, {\cal G}_0-\frac{\delta^2\Omega_{B,0}^2}{2\sigma^2}(\sinh^2{(s)}+\frac{\sinh^2{(2s)}}
 {\cosh{(4s)}}})\}.
\end{eqnarray}
This equation gives us a quantitative way to evaluate the contributions
of the curved background spacetime of the Earth to the steering for the $b_2\to b_1$  scenario, when the
satellites are far away to the Earth. It is clearly shown that ${\cal G}^{b_2\to b_1}$ is equal to ${\cal G}_0$  when the $\delta\rightarrow0$, which means that the effect induced by the
curved background spacetime of the Earth vanishes in this limit.

The  typical distance between the  ground station and the geostationary satellite is about $3.6\times10^4$km, which yields the height $r_B = 4.237\times10^4$ km for the satellite.
Since the height of current GPS (Global Position System) satellites is $r_B \approx 2.7\times10^4$km. For this distance the influence of relativistic disturbance of the spacetime curvature on quantum steerability cannot be ignored for the quantum information tasks at current level technology \cite{satellite1,satellite2,MJAG}. Hence, in this work the plotting range of  the satellite height will be constrained to geostationary satellites height.

In Fig. (3), we plot the quantum steerability ${\cal G}^{b_1\to b_2}$, as well as ${\cal G}^{b_2\to b_1}$ of the final state as a function of the height $h$. The plot range is limited to geostationary Earth orbits $r_B(GEO)=r_A+35784$ km. Here, the range of peak frequency parameter $\Omega_2$ is fixed from $0.6$ to $1$ to satisfy $\delta\ll(\frac{\delta\Omega_2}{\sigma})^2\ll1$. It is shown that both the $b_1\rightarrow b_2$ and $b_2\rightarrow b_1$ steering increase for a specific range of height parameter $h$ and then gradually approach to a finite value with  increasing $h$. This is because the
total frequency shift in Eq. (\ref{aw}) both includes the Schwarzschild term and the rotation
term. The parameter $\delta$ in the Kerr spacetime is $\delta=\frac{1}{8}\frac{r_S}{r_A}\big(\frac{1-2\frac{h}{r_A}}{1+\frac{h}{r_A}} \big)$ which is different from the Schwarzschild case $\delta_{Sch}^{'}=-\frac{r_S}{4r_A}\frac{h}{(r_A+h)}$ \cite{DEBT, DEBA} since special relativistic effects are involved \cite{kerr}. When the satellite moves at the height $h=\frac{r_A}{2}$, the Schwarzschild term $\delta_{Sch}$ vanishes and photons received on satellites will generate a very small frequency shift dominated by special relativistic effects, therefore the lowest order rotation term $\delta_{rot}$ needs to be considered. In addition, we can see that whatever ${\cal G}^{b_1\to b_2}$ or ${\cal G}^{b_2\to b_1}$ both reduce with increasing $h$ after reaches the peak. That's why we say the gravitational frequency shift effect is a lossy channel.  This losing degree of quantum steering depends on the dimensionless peak frequency of mode $b_2$, which means that this lossy channel not only depends on curvature of the Earth.

In fact, the peak value of quantum steering indicates the fact that the photon's frequency received by satellites experiences a transformation from blue-shift to red-shift, which causes the Gaussian steering between the photon pairs to increase first and then to reduce with increasing height  \cite{kerr}. In the Schwarzschild limit $a, \omega_A\to 0$, the frequency shift simplifies to
$\frac{\Omega_B}{\Omega_A}=\sqrt{\frac{1-\frac{2M}{r_A}}{1-\frac{3M}{r_B}}},$
from which we can see that  the  received photon's frequency
on satellites do not experience any frequency shift at  $h=\frac{r_A}{2}$. On the other hand, the frequency of photon received at orbits with height $h<\frac{r_A}{2}$ will experience  blue-shift, while the frequencies of photons received at height $h>\frac{r_A}{2}$ experience red-shift.
For this reason, the photons experience different frequency shifts when the satellite locates at different heights in the Kerr spacetime. Therefore, the Gaussian steering  increases at the beginning, reaches the peak value (corresponds to the satellite at the heights $h\approx\frac{r_A}{2}$, i.e. the parameter $\delta=0$ ), and then decreases with increasing height.

To check the asymmetric degree of steerability under the Earth's spacetime curvature,  we calculate the Gaussian steering asymmetry
\begin{equation}
{\cal G}^{\Delta}=| {\cal G}^{b_1\to b_2}-{\cal G}^{b_2\to b_1}|,
\end{equation}
and plot it as a function of the peak frequency $\Omega_2$ and the height $h$ of the satellite in Fig. (4). This allows us to have a better understanding of how the  peak frequency $\Omega_2$ and the  Earth's  gravitation affect steering asymmetry. It is shown that the ${\cal G}^{\Delta}$ is close to zero, i.e., the steerability is almost symmetric when the height parameter $h\to0$ and the peak frequency $\Omega_2\to0$ because ${\cal G}^{b_1\to b_2}\approx{\cal G}^{b_2\to b_1}$ in these two cases. In addition, the steering asymmetry monotonically increases with increasing orbit height $h$ of the satellite. The physical support behind this is that gravitational field would reduce quantum resource \cite {MAK}, and the effect of gravitational field on  different directions of steering is different \cite{steeringrqi2}. Furthermore, it is not difficult to infer that if the gravitation is strong enough or Bob is close to the horizon of a black hole, the gravitational frequency should lead to completely asymmetry: Alice can steer Bob but Bob cannot steer Alice at all.

\begin{figure}[tbp]
\centering
\includegraphics[height=2.1in, width=2.3in]{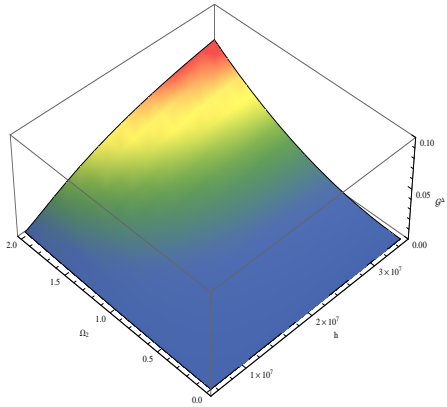}
\caption{(Color online) The Gaussian steering asymmetry ${\cal G}^{\Delta}$ as a function of the height $h$ of the satellite and the peak frequency $\Omega_2$ of mode $b_2$. The   Gaussian bandwidth and the squeezing parameter are fixed as $\sigma=1$ and $s=1$, respectively.
}\label{f3}
\end{figure}

\vspace*{0.8cm}

\section{Conclusions}

In conclusion, we have studied Gaussian steering for a two-mode Gaussian state when one of the modes propagates from the ground to satellites.
We found that the frequency shift induced by the curved spacetime  of the Earth reduces the quantum correlation of the steerability between the photon pairs  when  one of the entangled photons is sent to the Earth station and the other photon is sent  to the satellite.
In addition,  the influence of spacetime curvature on the steering in the
Kerr spacetime is very different from the non-rotation case because special relativistic effects are involved.
We also found that Gaussian steering is easier to change
with the initial squeezing parameter than the gravitational effect and other parameters. Although the gravitational effect of the Earth is small, it will lead to the Gaussian steering asymmetry between the photon pairs. This is because the influence of  gravitational field on the steering of the downlink setup  is stronger than the effect of gravitational field on the steering of the uplink setup, which results in the increase of quantum steering asymmetry. Therefore, we can conclude that the effects
induced by the curved spacetime of the Earth will generate quantum steering asymmetry. Finally, the peak value is found to be a critical point which indicates the received photons experience a transformation from  blue-shift to red-shift. According to the equivalence principle, the effects of acceleration are equivalence with the effects of gravity, our results could be in principle apply to dynamics of quantum steering under the influence of acceleration. Since realistic quantum systems will always exhibit gravitational
and relativistic features, our results should be significant both for giving more advices to realize quantum information protocols such
as quantum key distribution from Earth to satellites and for a general understanding of quantum steering in  relativistic quantum systems.

\begin{acknowledgments}
This work is supported by the Hunan Provincial Natural Science Foundation of China under Grant No. 2018JJ1016; and the National Natural Science Foundation
of China under Grant  No. 11675052, and No. 11475061.		
\end{acknowledgments}


\begin{thebibliography}{99}
\bibitem{E.P.C}
E. Schr\"odinger, \emph{Proc. Camb. Phil. Soc {\bf 1935},  31}, 553.

\bibitem{E.P.C1}
E. Schr\"odinger, \emph{Proc. Camb. Phil. Soc {\bf 1936}, 32}, 446.

\bibitem{CBEG}
C. Branciard, E. Cavalcanti, S. Walborn, V. Scarani, and
H. Wiseman, \emph{Phys. Rev. A {\bf 2012}, 85}, 010301.

\bibitem{HMW}
H. Wiseman, S. Jones, and A. Doherty, \emph{Phys. Rev. Lett {\bf2007},
98}, 140402.


\bibitem{Skrzypczyk}
P. Skrzypczyk, M. Navascu\'es, and D. Cavalcanti, \emph{Phys. Rev.
Lett {\bf 2014}, 112}, 180404.


\bibitem{Kocsis}
Q. He, Q. Gong, and M. Reid, \emph{Phys. Rev. Lett {\bf2015}, 114}
060402.

\bibitem{Adesso2015}
I. Kogias, A. Lee, S. Ragy, and G. Adesso, \emph{Phys. Rev. Lett {\bf 2015}, 114}, 060403.


\bibitem{steering3}
M. Marciniak, A. Rutkowski, Z. Yin, M. Horodecki, and R. Horodecki,
\emph{Phys. Rev. Lett {\bf 2015}, 115}, 170401.

\bibitem{steering4}
Q. He, L. Rosales-Z\'{a}rate, G. Adesso, and M. Reid,
\emph{Phys. Rev. Lett {\bf 2015}, 115}, 180502.

\bibitem{MWZZ}
M. Wang, Z. Qin, and X. Su, \emph{Phys. Rev. A {\bf 2017}, 95}, 052311.


\bibitem{steering5}
A. Sainz, N. Brunner, D. Cavalcanti, P. Skrzypczyk, and T. Vertesi,
\emph{Phys. Rev. Lett {\bf 2015}, 115}, 190403.

\bibitem{SLCC}
S. Chen, C. Budroni, Y. Liang, and Y. Chen
\emph{Phys. Rev. Lett {\bf 2016}, 116}, 240401.


\bibitem{Walborn}
S. Walborn, A. Salles, R. Gomes, F. Toscano, and P.
Souto-Ribeiro, \emph{Phys. Rev. Lett {\bf 2011}, 106}, 130402.


\bibitem{steering2}
C. Li, K. Chen, Y. Chen, Q. Zhang, Y. Chen, and J. Pan,
\emph{Phys. Rev. Lett {\bf 2015}, 115}, 010402.

\bibitem{Bowles}
J. Bowles, T. V\'ertesi, M. Quintino, and N. Brunner, \emph{Phys.
Rev. Lett {\bf 2014}, 112}, 200402.

\bibitem{VHTE}
V. H$\ddot{a}$ndchen, T. Eberle, S. Steinlechner, A. Samblowski, T. Franz, R. Werner and  R. Schnabel, \emph{Nat. Photon {\bf 2012}, 6}, 596.

\bibitem{VHTSS}
S. Wollmann, N. Walk, A. Bennet, H. Wiseman, and G. Pryde,
\emph{Phys. Rev. Lett {\bf 2016}, 116}, 160403.


\bibitem{Handchen}
T. Guerreiro, F. Monteiro, A. Martin, J. Brask, T. V\'{e}rtesi, B. Korzh, M. Caloz, F. Bussi$\grave{ e}$res, V. Verma, A. Lita, R. Mirin, S. Nam, F. Marsilli, M. Shaw, N. Gisin, N. Brunner, H. Zbinden, and R. Thew, \emph{Phys. Rev. Lett {\bf 2016}, 117}, 070404.


\bibitem{BWSR}
B. Wittmann, S. Ramelow, F. Steinlechner, N. Langford,
N. Brunner, H. Wiseman, R. Ursin, and A. Zeilinger,
\emph{New Journal of Physics {\bf 2012}, 14}, 053030.

\bibitem{SKMJ}
S. Kocsis, M. Hall, A. Bennet, and G. Pryde,
\emph{Nat. Commun {\bf 2015}, 6}, 5886.

\bibitem{DJSS}
D. Saunders, S. Jones, H. Wiseman, and G. Pryde,
\emph{Nat. Phys {\bf 2010}, 6}, 845.

\bibitem{TEVH}
T. Eberle, V. H$\ddot{a}$ndchen, J. Duhme, T. Franz, R. F-Werner, and R. Schnabel, \emph{Phys. Rev. A {\bf 2011}, 83}, 052329.

\bibitem{Xiao2017}
Y. Xiao, X. Ye, K. Sun, J. Xu, C. Li, and G. Guo,
\emph{Phys. Rev. Lett {\bf 2017}, 118}, 140404.

\bibitem{SWNW}
S. Wollmann, N. Walk, A. Bennet, H. Wiseman, and G. Pryde,
\emph{Phys. Rev. Lett {\bf 2016}, 116}, 160403.


\bibitem{steeringrqi}
M. Navascues, D. Perez-Garcia, \emph{Phys. Rev. Lett {\bf 2012}, 19}, 160405.

\bibitem{steeringrqi1}
C. Sab\'in, and G. Adesso, \emph{Phys. Rev. A  {\bf 2015}, 92}, 042107.

\bibitem{steeringrqi2}
J. Wang, H. Cao, J. Jing, and H. Fan, \emph{Phys. Rev. D {\bf 2016}, 93}, 125011.

\bibitem{steeringrqi3}
W. Sun, D. Wang, L. Ye, \emph{L. Phys. Lett {\bf 2017}, 14}, 9.


\bibitem{DEBT}
D. Bruschi, T. Ralph, I. Fuentes, T. Jennewein,
and M. Razavi, \emph{Phys. Rev. D {\bf 2014}, 90}, 045041.

\bibitem{DEBA}
D. Bruschi, A. Datta, R. Ursin, T. Ralph, and I. Fuentes,
\emph{Phys. Rev. D {\bf 2014}, 90}, 124001.

\bibitem{satellite1}
G. Vallone, D. Bacco, D. Dequal, S. Gaiarin, V. Luceri, G. Bianco, and P. Villoresi, \emph{Phys. Rev. Lett {\bf 2015}, 115}, 040502.

\bibitem{satellite2}
J. Yin, Y. Cao, Y. Li, S. Liao, L. Zhang, J. Ren, W. Cai, W. Liu, B. Li, H. Dai, G. Li, Q.  L, Y. Gong, Y. Xu, S. Li, F. Li, Y. Yin, Z. Jiang, M. Li, J. Jia, G. Ren, D. He, Y. Zhou, X. Zhang, N. Wang, X. Chang, Z. Zhu, N. Liu, Y. Chen, C. Lu, R. Shu, C. Peng, J. Wang, J. Pan,  \emph{Science  {\bf 2017}, 356}, 1140.

\bibitem{kerr}
J. Kohlrus, D. Bruschi, J. Louko,  and I. Fuentes,
\emph{EPJ Quantum Technology {\bf 2017}, 4}, 7.

\bibitem{Alclock}
C. Chou, D. Hume, T. Rosenband,
and D. Wineland,  \emph{Science  {\bf 2010}, 329}, 1630.


\bibitem{Zych}
M. Zych, F. Costa, I. Pikovski, and C. Brukner, \emph{Nat. Commun {\bf 2011}, 2}, 505.

\bibitem{MADE}
M. Ahmadi, D. Bruschi, and I. Fuentes, \emph{Phys. Rev. D {\bf 2014}, 89}, 065028.

\bibitem{MADE2}
M. Ahmadi, D. Bruschi, C. Sab\'in, G. Adesso, and
I. Fuentes, \emph{Sci. Rep {\bf 2014}, 4}, 4996.

\bibitem{SPK}
S. Kish and T. Ralph, \emph{Phys. Rev D {\bf 2016}, 93}, 105013.

%
%

\bibitem{wangsyn}
J. Wang, Z. Tian, J. Jing, and H. Fan,  \emph{Phys. Rev. D {\bf 2016}, 93}, 065008.

\bibitem{RQI8}
M. Fink, A. Rodriguez-Aramendia, J. Handsteiner, A. Ziarkash, F. Steinlechner, T. Scheidl, I. Fuentes, J. Pienaar, T, Ralph, and R. Ursin, \emph{Nat. Commun  {\bf2017}, 8}, 15304.

\bibitem{MANI}
M. Nielsen and I. Chuang,\emph{
Quantum computation and quantum information}(Cambridge
University Press, 2000).

\bibitem{Visser}
M. Visser, {\bf 2007}, arXiv:0706.0622.




\bibitem{RMWG}
R. Wald, \emph{General relativity} (The University of
Chicago Press, Chicago and London, 1984).

\bibitem{ULMQ}
U. Leonhardt, \emph{Measuring the Quantum State of Light,
Cambridge Studies in Modern Optics} (Cambridge University
Press, Cambridge, 2005).


\bibitem{TGDT}
T. Downes, T. Ralph, and N. Walk, \emph{Phys. Rev. A
{\bf 2013}, 87}, 012327.

\bibitem{PPRW}
P. Rohde, W. Mauerer, and C. Silberhorn, \emph{New Journal
of Physics {\bf 2007}, 9}, 91.

\bibitem{IKAR}
I. Kogias, A. Lee, S. Ragy, and G. Adesso,
\emph{Phys. Rev. Lett {\bf 2015}, 114}, 060403.

\bibitem{NCMS}
M. Razavi and J. Shapiro, \emph{Phys. Rev. A {\bf2006}, 73},
042303.

\bibitem{DNMP}
D. Matsukevich, P. Maunz, D. Moehring, S. Olmschenk,
and C. Monroe, \emph{Phys. Rev. Lett. {\bf 2008}, 100},
150404.


\bibitem{HMWS}
H. Wiseman, S. Jones, and A. Doherty, \emph{Phys. Rev. Lett
{\bf 2007}, 98}, 140402.

\bibitem{renyi}
G. Adesso, D. Girolami, and A. Serafini, \emph{Phys. Rev. Lett {\bf 2012}, 109},
190502.

\bibitem{MJAG}
M. Jofre, A. Gardelein, G. Anzolin, W. Amaya, J. Capmany,
R. Ursin, L. Penate, D. Lopez, J. Juan, J.
Carrasco, \emph{Opt. Express {\bf 2011}, 19}, 3825.

\bibitem{MAK}
M. Ahmadi, K. Lorek, A. Ch\c{e}ci\'{n}ska, H. Smith, R. Mann, A. Dragan,
\emph{Phys. Rev. D {\bf 2016}, 93}, 124031.

\end{thebibliography}
\end{document}